\documentstyle[pra,preprint,aps]{revtex}

\begin{document}
\bibliographystyle{prsty}
\title{Direct measurement of the fine-structure interval in 
alkali atoms using diode lasers}
\author{Ayan Banerjee, Umakant D. Rapol, and Vasant 
Natarajan\cite{email}}
\address{Department of Physics, Indian Institute of Science, 
Bangalore 560 012, INDIA}

\maketitle

\begin{abstract}
We demonstrate a technique for directly measuring the 
fine-structure interval in alkali atoms using two 
frequency-stabilized diode lasers. Each laser has a 
linewidth of order 1 MHz and precise tunability: one laser 
is tuned to a hyperfine transition in the $D_1$ line, and 
the other laser to a hyperfine transition in the $D_2$ line. 
The outputs of the lasers are fed into a scanning Michelson 
interferometer that measures the ratio of their wavelengths 
accurately. To illustrate the technique, we measure the 
fine-structure interval in Rb, and obtain a value of 
237.6000(3)(5) cm$^{-1}$ for the hyperfine-free $5P_{3/2} - 
5P_{1/2}$ interval. 
\end{abstract}
\pacs{32.10.Fn,42.62.Fi,42.55.Px}

The advent of semiconductor diode lasers has revolutionized 
atomic physics and particularly the field of laser 
spectroscopy \cite{WIH91}. Single-mode laser diodes with cw 
output powers of few 10s of mW are now available at very low 
cost. By placing the diodes in an external cavity and using 
optical feedback from an angle-tuned grating \cite{MSW92}, 
they can be made to operate at single frequency (single 
longitudinal mode) with linewidths of order 1 MHz and 
tunability over several nm. This has increased the access of 
researchers to atomic physics experiments which have earlier 
required expensive ring-cavity dye or Ti-sapphire lasers. 
For example, the field of laser cooling \cite{nobel97} has 
blossomed in the past decade as several alkali atoms and 
alkali-like ions have cooling transitions in the infrared 
which are accessible with stabilized diode lasers. The main 
advantages of diode lasers are in terms of their tunability 
and narrow spectral width. In addition, techniques such as 
saturated-absorption spectroscopy using 
counter-propagating pump and probe beams to eliminate the 
first-order Doppler effect can help resolve narrow hyperfine 
transitions within a given atomic line. It is thus possible 
to get an absolute frequency calibration of the laser by 
locking to such a transition. 

In this letter, we show that this well-established 
capability of diode lasers can be used in an important 
application: to make precise measurements of fine-structure 
intervals in alkali atoms. Knowledge of fine-structure 
intervals is important for several reasons, {\it e.g.\ }in 
the study of atomic collisions, astrophysical processes, and 
relativistic calculations of atomic energy levels. Alkali 
atoms are particularly important because of their widespread 
use in ultra-cold collision studies, photoassociation 
spectroscopy, atomic tests of parity violation, and more 
recently in Bose-Einstein condensation. The species we have 
studied is Rb, but the technique is more general and should 
prove useful in other alkali atoms (such as Li, K, and Cs), 
alkali-like ions, and indeed any system where the 
transitions are accessible with diode lasers. Our technique 
uses two diode lasers, one of which is tuned to the $D_1$ 
line and the other to the $D_2$ line. The output of the 
lasers is fed into a scanning Michelson interferometer to 
directly obtain their wavelength ratio. Using tabulated 
values \cite{MOO71} for the wavelength of the $D_2$ line, we 
extract the $D_1 - D_2$ splitting. The value we obtain for 
the $5P$ state of Rb has a precision of 5 parts in $10^8$, 
and can be potentially improved by another order of magnitude.

The interferometer for the ratio measurement, shown 
schematically in Fig.\ \ref{schematic}, has been described 
extensively in a previous publication \cite{BRW01} where we 
highlighted its use as a precision wavelength meter. For 
consistency with the terminology there, we will call the two 
lasers as ``reference'' and ``unknown'', respectively. The 
basic idea is to obtain the ratio of the two laser 
wavelengths using a scanning Michelson interferometer where 
both lasers traverse essentially the same path. As the 
interferometer is scanned, the interference pattern on the 
detectors goes alternately through bright and dark fringes 
and the ratio of the number of fringes counted is the ratio 
of the two wavelengths. The ratio 
obtained is a wavelength ratio in air, however, the 
wavelength ratio in vacuum (or equivalent frequency ratio) 
is easily calculated by making a small correction for the 
dispersion of air \cite{EDL66} between the reference 
wavelength and the unknown.

The fine-structure measurements are done with two diode 
laser systems. 
The first diode laser system (or ``reference'' laser) is 
built around a commercial single-mode diode (Mitsubishi 
ML60125R-01) with a nominal operating wavelength of 785 nm. 
The light is collimated using 
an aspheric lens to give an elliptic beam of 5.8 mm $\times$ 
1.8 mm $1/e^2$ diameter. The laser is frequency stabilized 
in a standard external-cavity design (Littrow configuration) 
using optical feedback from a 1800 lines/mm diffraction 
grating mounted on a piezoelectric transducer \cite{MSW92}. 
Using a combination of temperature and current control, the 
laser is tuned close to the 780 nm $D_2$ line in atomic Rb 
($5S_{1/2} \leftrightarrow 5P_{3/2}$ transition). A part of 
the output beam is tapped for Doppler-free 
saturated-absorption spectroscopy in a Rb vapor cell. The 
various hyperfine transitions (and crossover resonances) in 
the two common isotopes of Rb, $^{85}$Rb and $^{87}$Rb, are 
clearly resolved, as shown in the insets of Fig.\ 
\ref{rblevels}. The linewidth of the hyperfine peaks is 
15--20 MHz; this is somewhat larger than the 6.1 MHz natural 
linewidth and is primarily limited by power broadening due 
to the pump beam \cite{foot5}. The injection current into 
the laser diode is modulated slightly to obtain an error 
signal so that the laser can be locked to any of the peaks 
in the saturated-absorption spectrum. The second diode laser 
system (``unknown'' laser) is identical to the first one, 
except that the laser diode (SDL 5311-G1) has a nominal 
operating wavelength of 792 nm. After stabilization, it is 
tuned to the 795 nm $D_1$ line in atomic Rb ($5S_{1/2} 
\leftrightarrow 5P_{1/2}$ transition). The elliptic beams 
from the two lasers are fed into the Michelson 
interferometer. The large Rayleigh ranges ($\sim$34 m and 
$\sim$3 m in the two directions) ensure that the beams 
remain collimated over the length of the interferometer and 
diffraction effects are not significant.

For the fine-structure measurements, we lock the reference 
laser to the $5S_{1/2}, F=3 \leftrightarrow 5P_{3/2}, 
F'=(3,4)$ crossover resonance in $^{85}$Rb, i.e. midway 
between the $F=3 \leftrightarrow F'=3$ and the $F=3 
\leftrightarrow F'=4$ transitions \cite{foot2}. From the Rb   
energy-level tables \cite{MOO71} and measured hyperfine 
shifts \cite{AIV77}, this corresponds to a frequency of 
$3.8422958 \times 10^{14}$ Hz. To get several independent 
measurements of the 
fine-structure splitting, we lock the second laser to 
different hyperfine transitions of the $D_1$ line in the two 
isotopes of Rb. In each case, the wavelength ratio is 
measured about 100 times. The values are plotted as a 
histogram and a Gaussian fit to the histogram yields the 
mean ratio and the $1\sigma$ (statistical) error in the 
mean. The mean wavelength ratios obtained from four 
independent measurements are listed in Table \ref{ratios}. 
We extract the hyperfine-free 
fine-structure interval in vacuum by making two corrections 
to the measured ratio. First, we convert the wavelength 
ratio in air to a frequency ratio using the refractive index 
of air at the two wavelengths from Edl\'en's formula 
\cite{EDL66}: $n=1.000275163$ at 780 nm ($D_2$) and 
$n=1.000275068$ at 795 nm ($D_1$). Then we remove the 
hyperfine frequency shifts shown in Fig.\ \ref{rblevels}, 
which are known to 
sub-MHz accuracy \cite{AIV77}. The extracted values of the 
fine-structure interval are listed in the table, from this 
we obtain an average value of:
\[ 237.6000(3)(5) {\rm \ \ cm^{-1},} \]
with the quoted errors being statistical and systematic, 
respectively. Isotope shifts in the fine-structure interval
are negligible at this level of precision.

The main source of statistical error is that the frequency 
counter counts zero crossings and does not count fractional 
fringes. The total number of fringes counted depends on the 
fringe rate (or cart speed) coupled with the 10 s 
integration time. Currently our photodiode electronics 
limits the cart speed so that we can use only about 20 cm of 
cart travel per measurement. This results in a single-shot 
statistical error of about 5 parts in $10^7$ in each data 
set \cite{foot3}. The error of less than 5 parts in $10^8$ 
in Table \ref{ratios} comes after averaging over $\sim 100$ 
individual measurements. Since the coherence length of the 
stabilized diode laser is about 50 m, it should be possible 
to use at least 1 m of cart travel for each measurement with 
simple improvements in the counting electronics. This should 
help in reducing the statistical error in future to below 1 
MHz.

There are several potential sources of systematic error, the 
main two being variation in the lock point of the lasers and 
non-parallelism of the two laser beams in the 
interferometer. We have checked for the first error by 
measuring the fine-structure interval with the unknown laser 
on different hyperfine transitions of the $D_1$ line. As 
seen from Table \ref{ratios}, within the errors quoted, the 
different ratios yield consistent values for the 
hyperfine-free fine-structure interval. This implies that 
there is no significant variation in the lock point of the 
lasers. The second source of systematic error, namely that 
the two beams have a small angle between them, is more 
serious. Any misalignment would cause a systematic increase 
in the measured ratio given by $1/ \cos \theta$, where 
$\theta$ is the angle between the beams. We have tried to 
minimize this error in two ways. The first method is to use 
the unused output beam of the reference laser (the one on 
the same side of the beamsplitter as the input beam) as a 
tracer for aligning the unknown laser beam, and checking for 
parallelism over a distance of about 2 m \cite{foot4}. The 
second method is to check for parallelism by looking for a 
minimum in the measured ratio as the angle of the unknown 
beam is varied. This works because the measured value is 
always larger than the correct value, whether $\theta$ is 
positive or negative, and becomes minimum when $\theta = 0$. 

To get an estimate of the systematic error using these 
methods of alignment, we have measured the frequency ratio 
of two identical stabilized diode lasers locked to different 
hyperfine transitions in the $D_2$ line of $^{85}$Rb \cite{foot6}, 
as described 
in our earlier work \cite{BRW01}. The frequency difference 
is known to be 2944 MHz, while the measurements 
yield a mean value of 2932(16) MHz. This shows that the 
systematic error is below 16 MHz or about 0.0005 cm$^{-1}$. 
In future, this error can also be brought down to the 
sub-MHz level by carefully checking for parallelism over 
longer distances.

The value for the $5P$ fine-structure interval we obtain can 
be compared to the value listed in the Rb energy-level 
tables published by NIST \cite{MOO71}: 237.60 cm$^{-1}$. Our 
result is consistent with this value but has significantly 
higher accuracy. The NIST value is obtained by fitting the 
energy levels to all available spectroscopic data. 
The accuracy of 0.01 cm$^{-1}$ for this 
value is typical for most alkali atoms and our 
technique has the potential to improve this by two orders of 
magnitude. In general, fine-structure intervals are known 
very poorly compared to hyperfine shifts. This is because 
hyperfine shifts, which are of order GHz, are more 
accessible to techniques such as microwave resonance or 
heterodyne measurements \cite{YSJ96}, while fine-structure 
splittings are of order THz and cannot be measured very 
easily. It is in this range of frequency differences that 
our technique is uniquely suitable. Finally, we mention that 
the scanning interferometer can also be used as a precision 
wavelength meter, giving the unknown wavelength in terms of 
the reference \cite{BRW01}. This could be important, for 
example, in laser cooling of ions where it is necessary to 
set the laser frequency with $\sim$MHz precision but not 
always possible to find transitions to which the cooling 
laser can be locked. Recently, the absolute frequency of a 
diode laser stabilized on the Rb $D_2$ line has been 
measured with sub-MHz accuracy \cite{YSJ96}. Our 
wavelength meter can thus act as a secondary 
reference to improve the absolute precision of energy-level 
tables.

In conclusion, we have demonstrated a technique to directly 
measure the fine-structure interval in alkali atoms. Our 
method uses two stabilized diode lasers, one locked to 
the $D_1$ line and the other to the $D_2$ line, and a 
scanning Michelson interferometer to obtain their wavelength 
ratio. As an illustration of the power of this technique, we 
measure the interval in the $5P$ state of Rb to be 
237.6000(3)(5) cm$^{-1}$, or 7 123 069(9)(15) MHz. The 
method is easily extendable to other systems in which 
transitions are accessible with diode lasers. In addition, 
the interferometer with a diode laser locked to an atomic 
transition can be used as a precision wavelength meter for 
tunable lasers \cite{BRW01} since the atomic transition acts 
as a convenient calibrated wavelength marker.

The authors are grateful to Dipankar Das and St\'ephane Berciaud 
for help with the measurements. This work was supported by 
research grants from the Board of Research in Nuclear 
Sciences (DAE), and the Department of Science and 
Technology, Government of India.

\begin{figure}
\caption{
Schematic of the scanning interferometer. The Michelson 
interferometer consists of a beamsplitter (BS), two end 
mirrors (M), and two retro-reflectors (R) mounted back 
to back on a movable cart. 
The ratio of the number of fringes counted in the two 
detectors for a given cart travel yields the ratio of the 
two wavelengths.
}
\label{schematic}
\end{figure}

\begin{figure}
\caption{
Rb energy levels. The figure shows the relevant energy 
levels of $^{85}$Rb and $^{87}$Rb in the ground $5S$ state 
and first excited $5P$ state. The various hyperfine levels 
are labeled with the value of the total angular momentum 
quantum number $F$, and the number on each level is the 
energy displacement (in MHz) from the unperturbed state. The 
two insets show probe transmission as our diode laser is 
scanned across the $D_2$ line in a Doppler-free saturated 
absorption spectrometer. The inset on the left is for 
transitions starting from the $F=3$ state in $^{85}$Rb, and 
the one on the right is for transitions starting from the 
$F=2$ state in $^{87}$Rb.
}
\label{rblevels}
\end{figure}

\begin{table}
\caption{ 
The table lists the measured wavelength ratios and $5P$ 
fine-structure interval. The reference laser was locked to a 
transition of the $D_2$ line in $^{85}$Rb, corresponding to 
a frequency of $3.8422958 \times 10^{14}$ Hz, while the 
second laser was tuned to various hyperfine transitions of 
the $D_1$ line in $^{85}$Rb and $^{87}$Rb, as listed. The 
hyperfine-free interval in vacuum was extracted by first 
converting the wavelength ratio to a frequency ratio and 
then removing the hyperfine shifts shown in Fig.\ 
\ref{rblevels}. The errors are statistical $1 \sigma$ 
deviations.
}
\begin{tabular}{ccc}
Measured transition & Wavelength ratio & $5P_{3/2} - 
5P_{1/2}$ \\
($D_1$ line)      &                & (cm$^{-1}$) \\
\tableline
\vspace*{-2mm}\\
$^{85}{\rm Rb},F=2 \rightarrow F'=3 $ & 1.018 880 35(3) & 
237.6001(4) \\
$^{85}{\rm Rb},F=3 \rightarrow F'=2 $ & 1.018 889 54(4) & 
237.6002(5) \\
$^{87}{\rm Rb},F=1 \rightarrow F'=2 $ & 1.018 873 18(5) & 
237.6002(6) \\
$^{87}{\rm Rb},F=2 \rightarrow F'=1 $ & 1.018 893 79(3) & 
237.5995(4) \\
\end{tabular}
\label{ratios}
\end{table}


\begin{references}
\bibitem{foot5}
Collisional broadening at the gas pressure in the Rb vapor 
cell is estimated to be $\sim$10 kHz and the linewidth of 
the laser has been independently measured to be $\sim$1 MHz.
\bibitem{foot2}
We lock to crossover resonances because they are more 
prominent than the hyperfine peaks.
\bibitem{foot3}
The calculated error from missing one fringe over a distance 
of 20 cm is about 1 part in $10^6$ at 780 nm. Our error is 
smaller because we are measuring a wavelength ratio close to 
1 and chances are high that either both counters see a zero 
crossing or miss it.
\bibitem{foot4}
Ideally, the unknown beam should enter exactly where the 
reference beam exits so that both beams traverse exactly the 
same path, albeit in opposite directions. However, this 
causes one laser to feed into the other and destabilize it. 
Therefore we align the beams with a small offset and check 
for parallelism over a finite distance.
\bibitem{foot6}
The reference laser was locked to the $F=3 
\leftrightarrow F'=(3,4)$ crossover resonance. 
The unknown laser was locked to the $F=2 
\leftrightarrow F'=(2,3)$ crossover resonance, whose frequency 
is 2944 MHz higher (see Fig.\ 2).
\end{references}
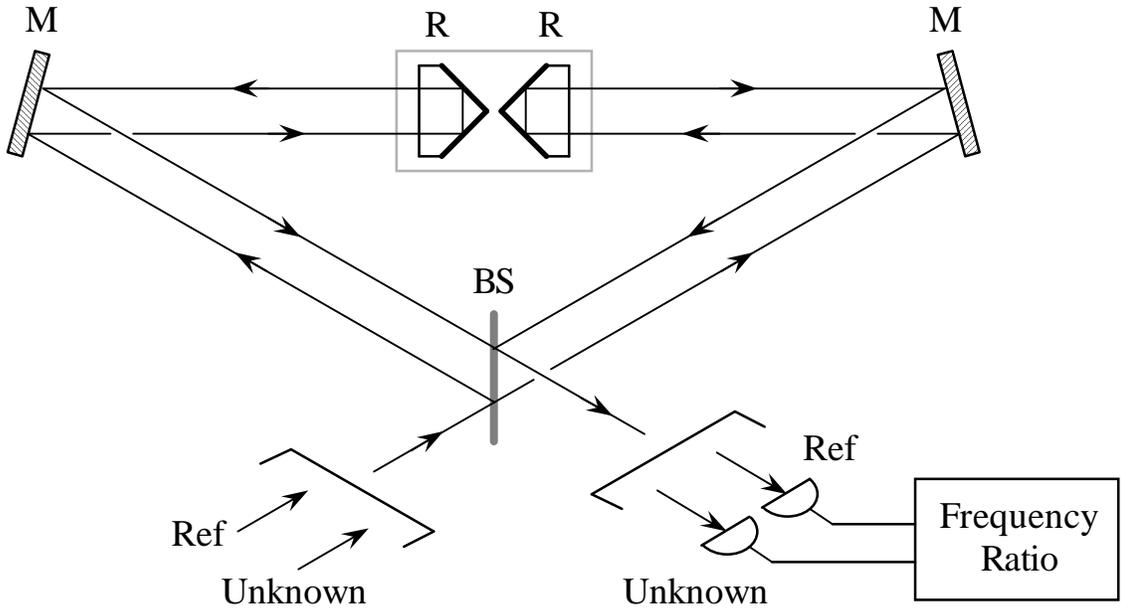
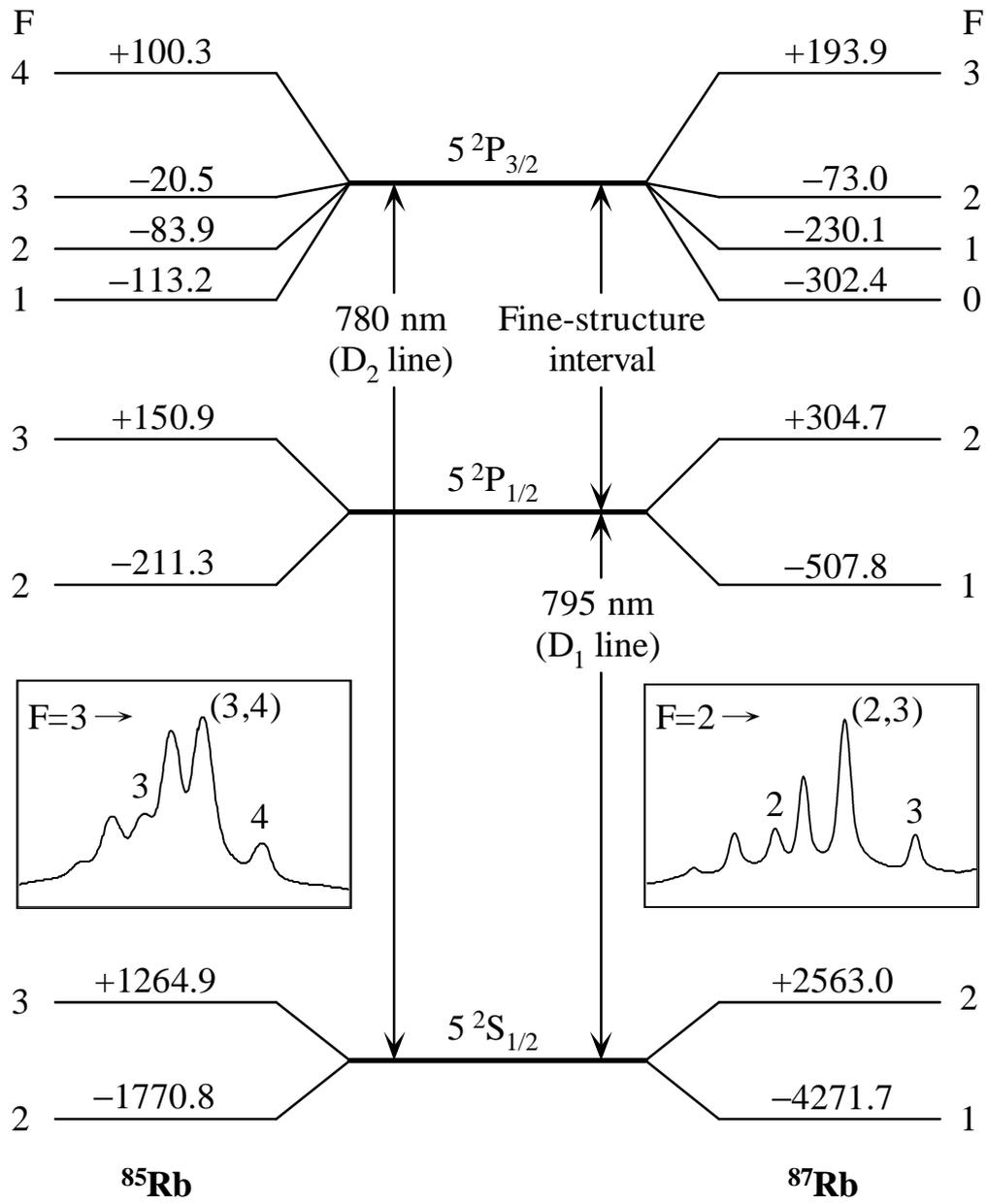
\end{document}